\documentclass{article}
\usepackage[utf8]{inputenc}

\usepackage[margin=1in]{geometry}
\usepackage{amsmath,amssymb}
\usepackage[natbibapa]{apacite}
\usepackage[colorlinks,citecolor=blue,urlcolor=blue]{hyperref}
\usepackage{graphicx}

\usepackage{fancyhdr}
\usepackage{xr}

\usepackage{caption}
\usepackage{subcaption}

\newcommand{\G}{\boldsymbol{G}}
\newcommand{\g}{\boldsymbol{g}}
\newcommand{\Hb}{\boldsymbol{H}}
\newcommand{\U}{\boldsymbol{U}}

\newcommand{\V}{\boldsymbol{V}}
\newcommand{\W}{\boldsymbol{W}}
\newcommand{\I}{\mathbb{I}}
\newcommand{\Pbb}{\mathbb{P}}

\makeatletter
\newcommand*{\addFileDependency}[1]{
  \typeout{(#1)}
  \@addtofilelist{#1}
  \IfFileExists{#1}{}{\typeout{No file #1.}}
}
\makeatother
\newcommand*{\myexternaldocument}[1]{%
    \externaldocument{#1}%
    \addFileDependency{#1.tex}%
    \addFileDependency{#1.aux}%
}
\myexternaldocument{supplementary} 

\begin{document}

\title{Statistical methods for Mendelian models \\ with multiple genes and cancers}

\author{Jane W. Liang$^{\ast, \dagger}$, Gregory E. Idos$^\ddagger$, Christine Hong$^\ddagger$, Stephen B. Gruber$^\ddagger$, \\ Giovanni Parmigiani$^\dagger$, and Danielle Braun$^\dagger$}
\date{\small \textit{$^\ast$Correspondence should be addressed to: jwliang@g.harvard.edu \\[4pt] $^\dagger$Department of Biostatistics, Harvard T.H. Chan School of Public Health, Boston, MA, USA \\ Department of Data Science, Dana-Farber Cancer Institute, Boston, MA, USA \\[4pt] $^\ddagger$Center for Precision Medicine, City of Hope, Duarte, CA, USA}}

\maketitle

\noindent We gratefully acknowledge support from the National Cancer Institute at the National Institutes of Health grants 5T32CA009337 (JWL) and 4P30CA006516 (GP).

\begin{abstract}
    Risk evaluation to identify individuals who are at greater risk of cancer as a result of heritable pathogenic variants is a valuable component of individualized clinical management. Using principles of Mendelian genetics, Bayesian probability theory, and variant-specific knowledge, Mendelian models derive the probability of carrying a pathogenic variant and developing cancer in the future, based on family history. Existing Mendelian models are widely employed, but are generally limited to specific genes and syndromes. However, the upsurge of multi-gene panel germline testing has spurred the discovery of many new gene-cancer associations that are not presently accounted for in these models. We have developed PanelPRO, a flexible, efficient Mendelian risk prediction framework that can incorporate an arbitrary number of genes and cancers, overcoming the computational challenges that arise because of the increased model complexity. We implement an eleven-gene, eleven-cancer model, the largest Mendelian model created thus far, based on this framework. Using simulations and a clinical cohort with germline panel testing data, we evaluate model performance, validate the reverse-compatibility of our approach with existing Mendelian models, and illustrate its usage. Our implementation is freely available for research use in the PanelPRO R package. \\[4pt] \textit{Keywords: Mendelian models, risk prediction, germline panel gene testing, precision prevention, pathogenic variants.}
\end{abstract}

\section{Introduction}
\label{sec:introduction}
Certain genetic variants are known to greatly increase one's risk for developing cancer \citep{foulkes2008inherited}. Estimating the probability of carrying cancer-causing variants is important for personalized clinical management \citep{win2013criteria}. Mendelian models use principles of Mendelian genetics, probability theory, and gene-specific knowledge to estimate the probability that an individual receiving counseling (the counselee) is a pathogenic variant carrier \citep{murphy1969application}. As input, they take a potentially detailed pedigree of family history, which contains information on the relationships between individuals in the counselee's family, as well as their ages, sexes, cancer diagnosis statuses, and ages at diagnosis. As training, they rely on external estimates of the allele frequencies of the cancer-causing variants by gene, as well as the sex-specific and age-dependent probabilities of developing cancer conditional on presence of a pathogenic variant (cancer penetrances). These are population-level quantities and can be estimated based on published literature. The model outputs the counselee's carrier probabilities conditional on family history and the future risk of developing cancer. 

Mendelian models have been widely adopted \citep{euhus2001understanding, riley2012essential}, but existing models are generally limited to a small number of genes and cancers within an individual syndrome. For example, Mendelian models implemented in the free, open-source BayesMendel R package \citep{chen2004bayesmendel} include BRCAPRO, which computes carrier probabilities for pathogenic variants on BRCA1 and BRCA2 using family history of breast and ovarian cancers \citep{berry1997probability, parmigiani1998determining}; MMRpro, which considers Lynch syndrome genes MLH1, MSH2, and MSH6 using family history of colorectal and endometrial cancers \citep{chen2006prediction}; Pancpro, which considers an unspecified susceptibility gene for pancreatic cancer \citep{wang2007pancpro}; and Melapro, which computes the carrier probability for a variant on CDKN2A using family history of melanoma \citep{wang2010estimating}. These models have been validated in the literature \citep{berry2002brcapro, chen2006prediction, wang2007pancpro, wang2010estimating}. Other Mendelian risk prediction models for breast and ovarian cancer susceptibility include Claus \citep{claus1994autosomal}, IBIS (Tyrer-Cuzick) \citep{tyrer2004breast}, and BOADICEA \citep{antoniou2008boadicea}. In the context of future risk, non-Mendelian models are also prevalent; many summarize family history into binary or categorical predictors, followed by fitting logistic regression or another statistical learning model \citep{couch1997brca1, frank2002clinical, gail1989projecting, shattuck1997brca1, vahteristo2001probability}. 

In recent years, many deleterious gene variants have been linked with the increased risk of various cancers. For example, ATM, CHEK2, and PALB2 \citep{antoniou2014breast, marabelli2016penetrance, schmidt2016age} are among those associated with higher risk of breast cancer, in addition to the previously established BRCA1 and BRCA2. Furthermore, there is rising evidence that syndromes once thought to be distinct are determined by variants that increase the risk of multiple cancers \citep{kastrinos2009risk, kastrinos2011inherited, hruban2010update, moran2012risk, kote2011brca2}. These discoveries point to the need for flexible, scalable models that account for these relationships, many of which may be known but not yet accounted for by existing models. Simultaneously, the decreasing cost of DNA sequencing \citep{plichta2016s} positions us to leverage data from multi-gene panel germline testing. As more associations are discovered using panel studies, these associations will also need to be effectively incorporated into models for quantitative risk assessment. 

The development of a multi-syndrome model necessitates extending the Mendelian model framework from syndrome-specific to a fully generalizable framework that can account for associations between any number of genes and cancers. The increased model complexity implied by the inclusion of many genes and cancers also requires overcoming significant computational challenges. In particular, the Elston-Stewart peeling algorithm \citep{elston1971general} used to compute the carrier probabilities involves summing over all possible genotype configurations of a pedigree, a burden that is exponential with respect to the number of genes in the model. This computational cost must be appropriately addressed in order for larger, complicated models to return results in a reasonable runtime.  

We present PanelPRO, a flexible, computationally efficient Mendelian risk model that incorporates an arbitrary number of genes and cancers. Generalizing beyond syndrome-specific approaches, we model carrier probabilities and cancer future risk probabilities for multi-syndrome associations for a large number of genes. For example, BRCAPRO identifies individuals at high risk for breast or ovarian cancers due to pathogenic BRCA1 and BRCA2 variants. PanelPRO can also incorporate penetrances of breast cancer specific to pathogenic variants of ATM, CHEK2, and PALB2, as well as return carrier probabilities for these genes. We note that the BOADICEA model \citep{lee2019boadicea} has recently been extended to account for information from all five of these genes, as well as the effects of polygenic risk scores and other risk factors. Additionally, BRCA1 and BRCA2 have been linked to other syndromes, including pancreatic cancer \citep{greer2007role}. Therefore, it may be beneficial to incorporate the penetrance of pancreatic cancer for BRCA1 and BRCA2 pathogenic variant carriers, allowing us to borrow strength across additional cancers and heighten our understanding of cross-syndrome effects. The framework introduced is highly flexible and can be used to obtain the future risk of cancer for all cancers in the model.

\cite{lee2021panelpro} presents the R software package implementation for PanelPRO. It covers the informatics aspects of PanelPRO; detailed statistical topics are not included in \cite{lee2021panelpro}, and are the focus of this work. Specifically, \cite{lee2021panelpro} presents the package workflow from a user standpoint, including information on how the PanelPRO package pre-processes user-specified pedigrees and imputes missing ages prior to model evaluation. It includes only synthetic example pedigrees, highlighting expected model inputs and illustrating package usage. It also provides computational benchmarking and some discussion of computational concerns. In contrast, here we investigate the methodological background on which the software is based, providing a formal statistical treatment, methodological details, and assumptions. We provide comprehensive notation for our statistical model, including the flexible incorporation of secondary cancers and additional risk modifiers, and derive conditions for model collapsibility. As previously mentioned in \cite{lee2021panelpro}, we reduce the computational expense of the peeling algorithm to a feasible level for clinical use by making some reasonable simplifying assumptions \citep{madsen2018efficient}. We scrupulously verify the comparability of the prediction performance of PanelPRO against existing models in simulations and evaluate model performance using simulated data and a validation study using a high-risk multi-gene panel testing dataset from the USC-
Stanford Hereditary Cancer Panel (HCP) Testing Study.

PanelPRO generalizes beyond syndrome-specific approaches while resolving computational challenges that arise from including many associations. Although the proposed model is general, allowing users to enter an arbitrary number of genes and cancers, for this study, we illustrate the model using a proposed PanelPRO-5BC, a five-gene (ATM, BRCA1, BRCA2, CHEK2, and PALB2) breast and ovarian cancer model; and a proposed PanelPRO-11, an eleven-gene (ATM, BRCA1, BRCA2, CDKN2A, CHEK2, EPCAM, MSH1, MSH2, MLH6, PALB2, and PMS2), eleven-cancer (brain, breast, colorectal, endometrial, gastric, kidney, melanoma, ovarian, pancreatic, prostate, and small intestine) model. These PanelPRO models are applied to simulated data to illustrate the properties and usage of our approach. For validation, we apply the PanelPRO-5BC and PanelPRO-11 models to the high-risk multi-gene panel testing HCP dataset \citep{idos2019multicenter}. The PanelPRO R package is freely available for research purposes at \url{https://projects.iq.harvard.edu/bayesmendel/panelpro} \citep{lee2021panelpro}. Code to perform the analysis and 
generate the figures in this paper is available at \url{https://github.com/janewliang/PanelRePROducible} \citep{liang2022panelprorepo}. 

In Section~\ref{sec:statistical_methods}, we detail the statistical framework for PanelPRO. In Section~\ref{sec:data}, we describe the data used for specifying the input parameters and for validating our models. The results from the simulation studies and data application are reported in Sections~\ref{sec:simulations} and \ref{sec:data_application}, respectively. We conclude with a discussion in Section~\ref{sec:discussion}.

\section{Statistical Methods}
\label{sec:statistical_methods}
The contents of Sections~\ref{sec:proposed_model}-\ref{sec:computation} expand on and provide additional context for the framework described in \cite{lee2021panelpro}, including a generalization of the genotype vectors to multiple variants/mutation states, as well as notation for net future risk and further context for net vs. crude penetrance and future risk estimates. The following sections, Sections~\ref{sec:additional_modifiers}-\ref{sec:collapsibility}, contain entirely new elements that cover extensions for secondary cancers and other additional risk modifiers and describe conditions for model collapsibility.

\subsection{Proposed Model}
\label{sec:proposed_model}
Suppose that a given counselee has a family of size $I$, and let $\G_i = \left( G_{1i}, \dots, G_{Ki} \right)$ be the genotype vector for the $i$th family member, $i= 1, \dots, I$, where the counselee is indexed by 1. Each element $G_{ki}$ stores individual $i$'s carrier status for a pathogenic variant in the $k$th gene. Let $\mathcal{G}$ denote the set of all $2^K$ possible genotypes. In the simple case where only wild-type and heterozygous carriers need to be considered, $G_{ki}$ is binary. Under this simplest case of only wild-type and heterozygous states for each gene, in a two-gene model, $\G_i$ has length $K=2$ and there are $2^2 = 4$ possible genotypes (noncarrier for both genes, carrier of only the first pathogenic variant, carrier of only the second pathogenic variant, and carrier of pathogenic variants for both genes). More generally, additional states can be assumed: for instance, we can differentiate between carriers of variants on one allele vs. both alleles. Under this scenario, the three possible states would be encoded as $0, 1,$ or $2$, corresponding to $3^K$ possible genotypes. One can also assume the presence of additional states implied by having multiple variant types for a given gene.  

Each individual in the family also has a vector of phenotypes $\Hb_i = \left( \Hb_{1i}, \dots , \Hb_{Ri} \right)$, where $R$ is the arbitrary but fixed number of cancers or other phenotypes in the model. For individual $i$, let $T_{ri}$ be the age of diagnosis for cancer $r$, $C_i$ be their censoring age (current age or age of death), and $\delta_{ri} = \I(T_{ri} \leq C_i)$ be the binary indicator for the occurrence of cancer $r$ before the censoring age. Then the observed history of cancer $r$ is $\Hb_{ri} = (C_i, \delta_{ri})$ when $\delta_{ri} = 0$ and $\Hb_{ri} = (C_i, \delta_{ri}, T_{ri})$ when $\delta_{ri} = 1$. 

Denote the binary indicator of the $i$th individual being male as $U_i$. Finally, let  $\Hb = \left( \Hb_1 , \dots, \Hb_I \right)$ and $\U = \left( \U_1, \dots , \U_I \right)$ indicate the phenotypes and sexes of all family members, respectively.

Of interest is the estimation of the counselee's conditional probabilities $\Pbb(\G_1 = \g_1 | \Hb, \U)$ for all genotypes $\g_1 \in \mathcal{G}$. To obtain the posterior distribution, we apply Bayes' rule as presented by \cite{murphy1969application} and described by \cite{lange2003mathematical}: 
\begin{align}
    \label{eq:bayes}
    \Pbb(\G_1=\g_1 | \Hb, \U) \propto \Pbb(\G_1=\g_1) \Pbb(\Hb | \G_1=\g_1, \U).
\end{align}
After computing the right-hand side, the posterior probabilities must be normalized so that they sum to 1. $\Pbb(\G_1=\g_1)$ is the population-based frequency of the counselee's observed genotype $\g_1$, which we generally estimate from published literature. $\Pbb(\Hb | \G_1=\g_1, \U)$ is the conditional probability of the observed cancers for the entire pedigree given that the counselee has genotype $\g_1$. It can be computed by integrating over the entire set of possible family genotypes, using the law of total probability: 
\begin{align}
    \label{eq:totalprob}
    & \qquad \Pbb(\Hb | \G_1=\g_1, \mathbf{U}) \nonumber \\
    & = \sum_{\g_2, \dots, \g_I \in \mathcal{G}} \Pbb(\Hb | \G_1 = \g_1, \dots, \G_I = \g_I, \mathbf{U}) \Pbb(\G_2=\g_2, \dots, \G_I=\g_I | \G_1=\g_1) \nonumber \\
    & = \sum_{\g_2, \dots, \g_I \in \mathcal{G}} \left[ \prod_{i=1}^I \prod_{r=1}^R \Pbb(\Hb_{ri} | \G_i=\g_i, U_i) \right] \Pbb(\G_2=\g_2, \dots, \G_I=\g_I | \G_1=\g_1). 
\end{align}
In Eq.~\ref{eq:totalprob}, the joint genotype distribution conditional on the counselee's genotype $\Pbb(\G_2, ..., \G_I | \G_1)$ is derived using properties of Mendelian inheritance. One must have estimates of the population allele frequency of each variant and know the exact relationship between the counselee and each relative in the pedigree. We assume genotype independence across the founding family members, followed by conditionally independent genotypes in subsequent generations \citep{berry1997probability}. In order to model $\Pbb(\Hb | \G_1, ..., \G_I, \U) = \prod_{i=1}^I \prod_{r=1}^R \Pbb(\Hb_{ri} | \G_i, \U_i)$ in the integration step, we assumed that the relatives' phenotypes are independent conditional on genotype and sex. $\Pbb(\Hb_{ri} | \G_i, \U_i)$ is computed from the sex-specific penetrances. 

Two types of penetrances can be used. The net penetrance is the genotype-specific probability of developing a phenotype by a given age, \textit{in the absence} of death and other competing risks. The crude penetrance is the genotype-specific probability of developing a phenotype by a given age \textit{prior to} dying or developing other phenotypes. We typically use estimates reported in peer-reviewed literature, which are most often net. In such cases, we convert the estimates to crude when needed. Our implementation uses net penetrances $\Pbb \left(T_{ri} = t \middle| \G_i, U_i \right)$ to calculate the carrier probabilities. Explicitly,
\begin{align}
    \Pbb \left(\Hb_{ri} \middle| \G_i, U_i \right) =
    \begin{cases}
        1 - \sum_{s=1}^{C_i} \Pbb \left(T_{ri} = s \middle| \G_i, U_i \right) & \text{ if } \delta_{ri} = 0 \\
        \Pbb \left(T_{ri} = T_{ri}^{obs} \middle| \G_i, U_i \right) & \text{ if } \delta_{ri} = 1, 
    \end{cases}
\end{align}
where $T_{ri}$ is the random variable and $T_{ri}^{obs}$ is the observed cancer age. We assume that the censoring process and deaths by causes unrelated to a given cancer are independent of the time to cancer diagnosis. We further assume that censoring and death by other causes are non-informative with respect to carrier status. An in-depth discussion of the trade-offs involved in these modeling choices is provided by \cite{Katki2008}. 

Once these carrier probabilities have been obtained, they can be used to calculate the future risk of developing any of the cancers in the model. The future risk that the counselee develops the cancer indexed by $r$ can be computed using either the net or crude penetrances, and has a slightly different interpretation depending on which penetrance estimates are used. 

The counselee's $t_0$-year net future risk for a given cancer indexed by $r$ is
\begin{align}
    & \qquad \Pbb(T_{r1} \leq C_1 + t_0 | T_{r1} > C_1, \Hb, \U) \nonumber \\
    & = \sum_{\g_1} \Pbb(\G_1 = \g_1 | \Hb, \U) \Pbb(T_{r1} \leq C_1 + t_0 | T_{r1} > C_1, \G_1 = \g_1, U_1) \nonumber \\
    & = \sum_{\g_1} \Pbb(\G_1 = \g_1 | \Hb, \U) \cdot \frac{\Pbb( C_1 \leq T_{r1} \leq C_1 + t_0 | \G_1 = \g_1, U_1)}{\Pbb(T_{r1} > C_1 | \G_1 = \g_1, U_1)}. 
\end{align}
The numerator and denominator of the fractional term in the final expression can be computed from the net genotype-specific penetrances. This $t_0$-year risk prediction can be interpreted as the probability of developing cancer $r$ within $t_0$ years, conditional on being disease-free at the counselee's current age and assuming a hypothetical world where there is no death or other phenotypes. 

More realistically, one may consider the crude future risk, which can be interpreted as the probability of developing cancer $r$ within $t_0$ years, conditional on being disease-free and alive at the current age and assuming that the counselee survives all competing risks and death. For individual $i$, let $T_{d,ri}$ be their age of death from causes other than cancer $r$, $T_{ri}^* = \min(T_{ri}, T_{d,ri})$ be their age of first outcome (either cancer $r$ or death from causes other than cancer $r$), and $J_{ri} = \I(T_{ri}^* = T_{ri})$ be the binary indicator for developing the $r$th cancer. The counselee's crude risk can be calculated as
\begin{align}
    & \qquad \Pbb(T_{r1}^* \leq C_1 + t_0, J_{r1} = 1 | T_{r1}^* > C_1, \Hb, \U) \nonumber \\
    & = \sum_{\g_1} \Pbb(\G_1 = \g_1 | \Hb, \U) \Pbb(T_{r1}^* \leq C_1 + t_0, J_{r1} = 1 | T_{r1}^* > C_1, \G_1 = \g_1, U_1) \nonumber \\
    & = \sum_{\g_1} \Pbb(\G_1 = \g_1 | \Hb, \U) \nonumber \\ 
    & \qquad\quad \cdot \frac{\Pbb(C_1 \leq T_{r1}^* \leq C_1 + t_0, J_{r1} = 1 | \G_1 = \g_1, U_1)}{\Pbb(T_{r1}^* > C_1, J_{r1} = 1 | \G_1 = \g_1, U_1) - \Pbb(T_{r1}^* \leq C_1, J_{r1} = 0 | \G_1 = \g_1, U_1)}, 
\end{align}
The fractional term in the final expression is computed directly from the crude penetrances for cancer $r$ and the probability distribution of dying by other causes. The distribution of death by other causes is an additional parameter that can be estimated based on peer-reviewed literature. PanelPRO outputs crude future risk estimates by default. 

\subsection{Computation}
\label{sec:computation}
The integration step over the counselee genotypes in computing the probabilities in a Mendelian model is usually obtained by applying the Elston-Stewart peeling algorithm \citep{elston1971general}. The peeling algorithm is broadly used and generalizable to any number of heritable sites specified in the model. It is linear with respect to the number of relatives $I$ in the counselee's family, but sums over all possible genotypes in the pedigree. Therefore, computation time is exponential with respect to the number of genes $K$, making integration challenging for even a moderate number of genes. The peeling and paring algorithm \citep{madsen2018efficient} restricts the set of possible genotypes $\mathcal{G}$ to the set of genotypes $\mathcal{G}_M$ that contain up to $M$ deleterious variants. Then, one applies local integration over the ``pared'' set of genotypes $\mathcal{G}_M$ to approximate $\Pbb(\Hb | \G_1 = \g_1, \U)$ as 
\begin{align}
    & \sum_{\g_2, \dots, \g_I \in \mathcal{G}_M} \Pbb(\Hb | \G_1=\g_1, \dots, \G_I=\g_I, \U) \Pbb(\G_2=\g_2, \dots, \G_I=\g_I | \G_1=\g_1).
\end{align}
When $M \geq K$, the peeling-paring algorithm is equivalent to the peeling algorithm. 

For example, if the model has $K=11$ genes, each of which is fairly uncommon in the population, it would be very rare to observe an individual with all 11 variants, or even 9 or 10 variants. It is also possible that many genotypes with several such pathogenic variants may not be viable. Illustrating again with the scenario of two states for each gene, if $M=2$ and $K=11$, there are only $\sum_{m=0}^M \binom{K}{m} = 67$ admissible genotypes in $\mathcal{G}_M$ that need to be summed over. This is a substantial reduction compared to the $M^K = 2048$ in the original space $\mathcal{G}$. 

There are trade-offs in selecting the value for $M$. Higher values of $M$ lead to more precise approximations of the posterior genotypic distribution, but as $M$ approaches $K$, the computational complexity approaches $2^K$ once again. A smaller value of $M$ leads to higher efficiency, but this simplifying assumption has the cost of lower precision. These trade-offs are discussed in greater detail by \cite{madsen2018efficient}. Our PanelPRO implementation allows users to set the value for $M$, but in both our simulations and data application, we used $M=2$. Even as the number of genes in PanelPRO increases to 5, 7, 11, and potentially more, the computational expense of the integration step is kept feasible by imposing this restriction. The Lander-Green algorithm \citep{lander1987construction} is an alternative algorithm that is expected to be more efficient when the number of genes $K$ exceeds the number of family members $I$ in the pedigree. However, it may be less effective for large pedigrees. Further discussion of genetic linkage algorithms and their computational considerations can be found in \cite{lee2021panelpro}. 

\subsection{Additional Modifiers}
\label{sec:additional_modifiers}

\subsubsection{Secondary Cancers} 
Separate from the general approach of including primary cancers in a model, secondary cancers such as contralateral breast cancer modify risk estimation based on years since the primary cancer diagnosis. Suppose that primary breast cancer is indexed by $r = b$ and contralateral breast cancer by $r = c$. Then $\Pbb \left(\Hb_{ci} \middle| \G_i, U_i \right)$ can be expressed in terms of the contralateral breast cancer net penetrances $\Pbb \left(T_{ci} = t \middle| \G_i, U_i \right)$ as
\begin{align}
    \Pbb \left(\Hb_{ci} \middle| \G_i, U_i \right) =
    \begin{cases}
        1 - \sum_{s=T_{bi}^{obs}}^{C_i} \Pbb \left(T_{ci} = s \middle| \G_i, U_i \right) & \text{ if } \delta_{bi} = 1, \delta_{ci} = 0 \\
        \Pbb \left(T_{ci} = T_{ci}^{obs} - T_{bi}^{obs} \middle| \G_i, U_i \right) & \text{ if } \delta_{bi} = 1, \delta_{ci} = 1 \\
        1 & \text{ if } \delta_{bi} = 0. 
    \end{cases}
\end{align}
$T_{ci}$ is a random variable, whereas $T_{bi}^{obs}$ and $T_{ci}^{obs}$ are the observed ages of diagnosis for breast and contralateral breast cancer, respectively. 

\subsubsection{Risk Modifiers} 
Our implementation supports modifying risk at the cancer penetrance level using risk or hazard ratios for interventions such as prophylactic mastectomies, oophorectomies, and hysterectomies. When calculating carrier probabilities, simply use the modified penetrances $\Pbb^{\text{mod}} \left(T_{ri} \middle| \G_i, U_i \right)$ in place of the raw penetrances $\Pbb \left(T_{ri} \middle| \G_i, U_i \right)$. Suppose family member $i$ received a preventive intervention at age $T_{\text{int},i}$, and let $\text{RR}_{r,\text{int}} \left(T_{\text{int},i} \middle| \G_i, U_i \right)$ be the relative risk of developing cancer $r$ for those that receive the intervention at age $T_{\text{int},i}$, computed based on probabilities that are conditional on genotype and sex. Their modified penetrance for cancer $r$ is
\begin{align}
    \Pbb^{\text{mod}} \left(T_{ri} = t \middle| \G_i, U_i \right) & = \Pbb \left(T_{ri} = t \middle| \G_i, U_i \right) \nonumber \\ 
    & \qquad \cdot \left[ \I(t < T_{\text{int},i}) +  \text{RR}_{r,\text{int}} \left(T_{\text{int},i} \middle| \G_i, U_i \right) \I(t \geq T_{\text{int},i}) \right]. 
\end{align}

Suppose that instead of a relative risk, the hazard ratio for the intervention at age $T_{\text{int},i}$, $\text{HR}_{r,\text{int}} \left(T_{\text{int},i} \middle| \G_i, U_i \right)$, is available. Define $S_r \left(t \middle| \G_i, U_i \right) = 1 - \sum_{s=1}^t \Pbb \left(T_{ri} = s \middle| \G_i, U_i \right)$ as the survival function for cancer $r$ (conditional on sex and genotype), and let $\lambda_r \left(t \middle| \G_i, U_i \right) = \Pbb \left(T_{ri} = t \middle| \G_i, U_i \right)/S \left(t \middle| \G_i, U_i \right)$ be the corresponding hazard. Then the modified hazard, survival, and penetrance functions at age $t$ for this intervention are: 
\begin{align}
    \lambda_r^{\text{mod}} \left(t \middle| \G_i, U_i \right) & = \lambda_r \left(t \middle| \G_i, U_i \right) \left[ \I(t < T_{\text{int},i}) + \text{HR}_r(T_{\text{int},i}) \I(t \geq T_{\text{int},i}) \right] \nonumber \\
    S_r^{\text{mod}} \left(t \middle| \G_i, U_i \right) & = \prod_{s=1}^t (1 - \lambda_r^{\text{mod}} \left(s \middle| \G_i, U_i \right)) \nonumber \\
    \Pbb^{\text{mod}} \left(T_{ri} = t \middle| \G_i, U_i \right) & = S_r^{\text{mod}} \left(t + 1 \middle| \G_i, U_i \right) - S_r^{\text{mod}} \left(t \middle| \G_i, U_i \right).
\end{align}

\subsubsection{Germline and Tumor Biomarker Testing}
When germline or tumor marker testing results are available for some family members, this information can also be incorporated into the probability estimates. For each relative $i$, one substitutes $\Pbb(\Hb_{i} | \G_i, U_i) = \prod_{r=1}^R \Pbb(\Hb_{ri} | \G_i, U_i)$ with
\begin{align}
    \Pbb^{\text{mod}}(\Hb_{i} | \G_i, U_i) & = \Pbb(\Hb_{i} | \G_i, U_i) \cdot \Pbb (\V_i |\G_i) \cdot \Pbb (\W_i |\G_i), 
\end{align}
where $\V_i$ is the observed germline testing results and $\W_i$ is the observed tumor marker testing results for relative $i$. The probabilities $\Pbb (\W_i |\G_i)$ can be interpreted as test sensitivities and specificities. In practice, we often assume conditional independence of the different germline and marker testing results when robust joint estimates are not available. Currently, our implementation allows for modifying risk based on germline testing for any of the genes included in the model. It also supports modifying the risk of breast cancer conditional on BRCA1 or BRCA2 carrier status based on breast cancer tumor markers, ER, CK5/6, CK14, HER, and PR; and the risk of colorectal cancer conditional on MLH1, MSH2, MSH6, or PMS2 carrier status based on the colorectal cancer tumor marker, MSI. The implementation is designed to make it straightforward to extend risk modification to additional tumor markers or to modify the values according to user inputs.

\subsection{Conditions for Collapsibility}
\label{sec:collapsibility}

It is possible to collapse a larger ``full'' model into a smaller submodel that contains a subset of the specified genes and/or cancers. In this section we derive the following important facts: 

\begin{enumerate}
    \item If the submodel contains a subset of the genes in the full model, both models will return the same posterior probabilities if the allele frequency for being a noncarrier is 1 for all genes not present in the submodel. In other words, the probability of carrying a pathogenic variant is zero for any of the genes that are absent in the submodel.
    
    \item If the submodel contains a subset of the cancers in the full model, the submodel can be collapsed within the full model if the penetrances of the cancers not included in the submodel do not change depending on genotype. 
\end{enumerate}

Suppose that one has fit a model with $K$ genes and wishes to collapse it into a submodel that only includes the genes with indices in the set $\mathcal{K}^{(*)}$. Let $\mathcal{K}^{(-*)}$ be the index set for genes not included in the submodel, such that $\mathcal{K}^{(*)} \cup \mathcal{K}^{(-*)}$ is the full set of $K$ gene indices. Similarly, $\G_i$ can be partitioned as as $\G_i = \left( \G_i^{(*)}, \G_i^{(-*)} \right)$, where $\G_i^{(*)}$ is the genotype vector for the genes included in the submodel and $\G_i^{(-*)}$ is the genotype vector for the genes not included in the submodel. For the $i$th individual and $k$th gene, let $G_{ki} = 0$ represent a noncarrier and allow carriers of pathogenic variants to be represented using numbers other than zero. In order to obtain the same results from both the full model and submodel, a sufficient condition for collapsibility is that $\Pbb \left( G_{ki}^{(*)} = 0 \right) = 1$ for all genes $k \in \mathcal{K}^{(-*)}$ and all individuals $i$ in the family.

Under this condition, the counselee's genotype frequency is 
\begin{align}
    \Pbb(\G_1 = \g_1) & = \prod_{i=1}^K \Pbb(G_{k1} = g_{k1}) = \begin{cases}
        0 & \text{if } g_{k1} \neq 0 \text{ for at least one } k \in K^{(-*)} \\
        \Pbb \left(G_{k1}^{(*)} = g_{k1}^{(*)} \right) & \text{if } g_{k1} = 0 \text{ for all } k \in K^{(-*)}. 
    \end{cases}
\end{align}
Under the Mendelian model's conditional independence assumptions for the genotypes in the pedigree, we can also obtain
\begin{align}
    \Pbb(\G_2=\g_2, ..., \G_I=\g_I | \G_1=\g_1) = \Pbb\left( \G_2^{(*)}=\g_2^{(*)}, ..., \G_I^{(*)}=\g_I^{(*)} \middle| \G_1^{(*)}=\g_1^{(*)} \right)
\end{align}
when $g_{ki} = 0$ for all $k \in K^{(-*)}$ and $i = 1, \dots, I$; and $\Pbb(\G_2=\g_2, ..., \G_I=\g_I | \G_1=\g_1) = 0$ otherwise. 

Let $\mathcal{K}_{i,1}$ be the set of gene indices that satisfy $g_{ki} = 1$ for a given individual $i$, and let $\mathcal{K}_{i,1}^{(*)}$ be the analogous index set for only the genes that are included in the submodel. In the case where $g_{ki} = 0$ for all $k \in K^{(-*)}$, it follows that $\mathcal{K}_{i,1} = \mathcal{K}_{i,1}^{(*)}$ and
\begin{align}
    \Pbb(\Hb_{ri} | \G_i=\g_i, U_i) & = \prod_{k \in \mathcal{K}_{i,1}} \Pbb(\Hb_{ri} | G_{ki} = g_{ki}, U_i) \nonumber \\
    & = \prod_{k \in \mathcal{K}_{i,1}^{(*)}} \Pbb \left( \Hb_{ri} | G_{ki} = g_{ki}, U_i \right) \nonumber \\
    & = \Pbb \left(\Hb_{ri} \middle| \G_{i}^{(*)} = g_{i}^{(*)}, U_i \right). 
\end{align}

Using these findings to compute Eq.~\ref{eq:totalprob}, the expression for the posterior probabilities (Eq.~\ref{eq:bayes}) can be shown to be equivalent for the full model with $K$ genes and the submodel with $| \mathcal{K}^{(*)} |$ genes. Note that the equality below holds because the terms in the summation where $g_{ki} \neq 0$ for at least one $k \in \mathcal{K}^{(-*)}$ are zero, and therefore do not contribute. 
\begin{align}
    & \qquad \Pbb(\G_1=\g_1 | \Hb, \U) \nonumber \\
    & \propto \Pbb(\G_1=\g_1) \sum_{\g_2, \dots, \g_I \in \mathcal{G}} \left[ \prod_{i=1}^I \prod_{r=1}^R \Pbb(\Hb_{ri} | \G_i=\g_i, U_i) \right] \nonumber \\
    & \hspace{40mm} \cdot \Pbb(\G_2=\g_2, ..., \G_I=\g_I | \G_1=\g_1) \nonumber \\
    & = \Pbb \left( \G_1^{(*)}=\g_1^{(*)} \right) \sum_{\g_2^{(*)}, \dots, \g_I^{(*)} \in \mathcal{G}^{(*)}} \left[ \prod_{i=1}^I \prod_{r=1}^R \Pbb \left(\Hb_{ri} | \G_i^{(*)}=\g_i^{(*)}, U_i \right) \right] \nonumber \\
    & \hspace{54mm} \cdot \Pbb \left( \G_2^{(*)}=\g_2^{(*)}, ..., \G_I^{(*)}=\g_I^{(*)} \middle| \G_1^{(*)}=\g_1^{(*)} \right) \nonumber \\
    & \propto \Pbb\left( \G_1^{(*)}=\g_1^{(*)} \middle| \Hb, \U \right) 
\end{align}

Now suppose that the full model has $R$ cancers or other phenotypes and that the submodel of interest contains only the phenotypes with indices in the set $\mathcal{R}^{(*)}$. Denote the complementary set as $\mathcal{R}^{(-*)}$. Let $\Hb_{i}^{(*)}$ be the phenotype vector for individual $i$ corresponding to the phenotypes included in the submodel. 
In order to collapse the submodel within the full model, we require $\Pbb \left( \Hb_{ri} | \G_i = \g_i, U_i \right)$ to be constant for all $\g_i \in \mathcal{G}$ with respect to each individual $i$ and cancer $r \in \mathcal{R}^{(-*)}$. In words, each relative's penetrances for the cancers/phenotypes not included in the submodel must not depend on genotype. When this value is constant across genotypes, the additional cancers do not contribute to the posterior probabilities (Eq.~\ref{eq:bayes}) after normalization: 
\begin{align}
    & \qquad \Pbb(\G_1=\g_1 | \Hb, \U) \nonumber \\
    & \propto \Pbb(\G_1=\g_1) \sum_{\g_2, \dots, \g_I \in \mathcal{G}} \left[ \prod_{i=1}^I \prod_{r=1}^R \Pbb(\Hb_{ri} | \G_i=\g_i, U_i) \right] \nonumber \\
    & \hspace{40mm} \cdot \Pbb(\G_2=\g_2, ..., \G_I=\g_I | \G_1=\g_1) \nonumber \\
    & \propto \Pbb(\G_1=\g_1) \sum_{\g_2, \dots, \g_I \in \mathcal{G}} \left[ \prod_{i=1}^I \prod_{r \in \mathcal{R}^{(*)}} \Pbb(\Hb_{ri} | \G_i=\g_i, U_i) \prod_{r \in \mathcal{R}^{(-*)}} \Pbb(\Hb_{ri} | \G_i=\g_i, U_i) \right] \nonumber \\
    & \hspace{40mm} \cdot \Pbb(\G_2=\g_2, ..., \G_I=\g_I | \G_1=\g_1) \nonumber \\
    & = \Pbb(\G_1=\g_1) \left[ \prod_{i=1}^I \prod_{r \in \mathcal{R}^{(-*)}} \Pbb(\Hb_{ri} | \G_i=\g_i, U_i) \right] \nonumber \\ 
    & \qquad \cdot \sum_{\g_2, \dots, \g_I \in \mathcal{G}} \left[ \prod_{i=1}^I \prod_{r \in \mathcal{R}^{(*)}} \Pbb(\Hb_{ri} | \G_i=\g_i, U_i) \right] \Pbb(\G_2=\g_2, \dots, \G_I=\g_I | \G_1=\g_1) \nonumber \\
    & \propto \Pbb(\G_1=\g_1) \sum_{\g_2, \dots, \g_I \in \mathcal{G}} \left[ \prod_{i=1}^I \prod_{r \in \mathcal{R}^{(*)}} \Pbb(\Hb_{ri} | \G_i=\g_i, U_i) \right] \nonumber \\
    & \hspace{40mm} \cdot \Pbb(\G_2=\g_2, \dots, \G_I=\g_I | \G_1=\g_1) \nonumber \\
    & = \Pbb(\G_1=\g_1) \Pbb \left( \Hb^{(*)} \middle| \G_1=\g_1, \U \right) \nonumber \\
    & \propto \Pbb\left( \G_1=\g_1 \middle| \Hb^{(*)}, \U \right)
\end{align}

When the peeling-paring algorithm is used to restrict the possible genotypes to those in $\mathcal{G}_M$, one can simply substitute $\mathcal{G}_M$ for $\mathcal{G}$ in the derivations.

\section{Data}
\label{sec:data}
\subsection{Model Parameters}

In summary, the PanelPRO framework can be used to train a Mendelian model with any $K$ genes and $R$ cancers. The training involves meta-analysis or literature reviews to identify allele frequency inputs for each of the genes and age-dependent penetrance inputs for each of the gene-cancer combinations. PanelPRO converts these into clinically relevant carrier probabilities and absolute risk evaluations.

To illustrate, we present three different examples of PanelPRO models that collectively span 11 genes, plus a hypothetical PANC gene, and 11 cancers. PANC is an unspecified pancreatic cancer susceptibility gene used in the PancPRO \citep{wang2007pancpro} Mendelian model for identifying individuals at high risk for pancreatic cancer. The current PanelPRO package includes all the input estimates discussed here, as well as others, with the exception of PANC, which is only discussed here in the context of back-compatibility with PancPRO. Allele frequencies and penetrances used to construct the models are estimated based on existing literature. For BRCA1 and BRCA2, we used the non-Ashkenazi, Ashkenazi Jewish and Italian allele frequency estimates from BRCAPRO \citep{chen2004bayesmendel, antoniou2002comprehensive}; for MLH1, MSH2, and MSH6, we used the allele frequency estimates from MMRpro \citep{chen2004bayesmendel, chen2006prediction}; for the hypothetical PANC gene, we used the allele frequency estimate from Pancpro \citep{chen2004bayesmendel, wang2007pancpro}; and for CDKN2A, we used the allele frequency estimate from Melapro \citep{chen2004bayesmendel, berwick2006prevalence}. Allele frequency estimates for ATM, CHEK2, and PALB2 were taken from \cite{lee2016incorporating}. Finally, we estimated allele frequencies for EPCAM and PMS2 from a 25-gene panel study of 252,223 individuals \citep{rosenthal2017clinical}. 

Racially-informed (All Races, American Indian and Alaska Native, Asian, Black, White, Hispanic, White Hispanic, and White Non-Hispanic) probabilities of developing cancer (unconditional on genotype) were taken from the DevCan database \citep{statistical2020devcan}; penetrances for non-carriers are estimated by PanelPRO based on these values as well as the carrier penetrances and allele frequencies. When available, we took cancer penetrances for carriers from data included in the BayesMendel package: the BRCA1 and BRCA2 estimates for the probability of developing breast or ovarian cancer \citep{chen2020penetrance}; the MLH1, MSH2, and MSH6 estimates for the probability of developing colorectal or endometrial cancer \citep{wang2020penetrance, felton2007constitutive}; the PANC estimates for the probability of developing pancreatic cancer  \citep{wang2007pancpro, klein2002evidence}; and the CDKN2A estimates for the probability of developing melanoma \citep{wang2010estimating, begg2005lifetime, bishop2002geographical}. These gene$\times$cancer combinations are colored orange in Figure~1. Combinations colored in blue were obtained based on a literature review and are also available in the All Syndromes Known to Man Evaluator (ASK2ME) \citep{braun2018clinical} tool. Combinations in white are not included, as insufficient evidence of association has been accrued.

The sensitivities and specificities for tumor marker testing of ER, CK5/6, CK14, HER, and PR are from \cite{lakhani2002pathology, lakhani2005prediction}; the sensitivity and specificity for MSI testing are from \cite{chu2007random}. While the probability distribution of death from other causes does not come into play for the analysis in this paper, the PanelPRO package takes estimates for these values from DevCan \citep{statistical2020devcan}. The hazard ratios used for incorporating prophylactic mastectomies and oophorectomies into risk assessment are based on \cite{katki2007incorporating}. 

\subsection{USC-Stanford HCP Cohort}
\label{sec:usc}
In Section \ref{sec:data_application}, we validate our method using data from the USC-Stanford Hereditary Cancer Panel (HCP) Testing Study, a prospective, multi-center study on multiplex gene panel testing for cancer susceptibility \citep{idos2019multicenter}. A diverse group of 2000 patients was recruited from three medical centers: 1) USC Norris Comprehensive Cancer Center; 2) Los Angeles County + USC Medical Center; 3) Stanford University Cancer Institute. Eligible patients met clinical guideline criteria for genetic testing or had a $\geq$2.5\% probability of mutation carriage calculated by the following validated models or algorithms: BayesMendel \citep{chen2004bayesmendel}, BOADICEA \citep{antoniou2008boadicea}, IBIS (Tyrer-Cuzick) \citep{tyrer2004breast}, PREMM 1,2,6 \citep{euhus2001understanding, kastrinos2009risk, kote2011brca2}, National Comprehensive Cancer Network (NCCN) Guidelines, or a personal history of $\geq$10 cumulative lifetime tubular adenomas \citep{idos2019multicenter}. Diagnostic yield and off-target mutation detection was evaluated for 25- or 28-gene multi-gene panels \citep{idos2019multicenter} and performed by Myriad Genetic Laboratories (Salt Lake City, UT). 

We validate two models, PanelPRO-5BC and PanelPRO-11, using this data. We excluded families if: the counselee is a carrier of one or more variants only in genes not included in the model; the counselee has a variant of uncertain significance (VUS), but not a pathogenic variant for any of the genes in the model; or the models cannot be applied to the pedigree (typically due to ``loops''/inter-marriages or the reporting of cancer affection ages that are greater than the individual's censoring age). We thus consider 1612 families for validating PanelPRO-5BC (average family size of 34.15) and 1468 families for PanelPRO-11 (average family size of 33.78). Tables~1 and 2 summarize the number of pathogenic variant carriers and cancer cases among the counselees in the study for PanelPRO-5BC and PanelPRO-11, respectively. The differing number of carriers and cancers in Table~1 vs Table~2 is due to our pre-processing approach, which is partially dependent on the genes included in the model. 

Additionally, breast tumor marker and MSI testing results are reported for some of families. In the PanelPRO-5BC validation subset, there are 204 families with at least one relative tested for ER, 167 for PR, and 55 for HER (a total of 226 or 14\% families reported any results). In the PanelPRO-11 validation subset, there are 185 families with at least one relative tested for ER, 153 for PR, 45 for HER, and 3 for MSI (a total of 208 or 14.2\% families reported any results). Information on preventative interventions like mastectomies, oopherectomies, and hysterectomies was not available. 

We used racial or ancestry-informed allele frequencies and penetrances when available to match those reported in the data. Missing ages of cancer diagnosis and censoring ages were imputed by PanelPRO \citep{lee2021panelpro}.

\section{Simulation Studies}
\label{sec:simulations}
\subsection{Data Generation}
For each simulation study, we generated 1,000,000 families with structures sampled from families in the HCP study. Each simulated family includes the counselee and their grandparents, parents, and siblings, as well as any other offspring of the above. Genotypes for the founders of each family were sampled from non-Ashkenazi Jewish population-level allele frequencies. Then, the genotypes of their descendants were assigned based on Mendelian inheritance, with all genes assumed to be independent. The genotypes of non-blood relatives (i.e. spouses of the descendants) were sampled from the non-Ashkenazi Jewish population-level allele frequencies. Cancer status and age of diagnosis were simulated conditional on the individual's genotype, based on the cancer penetrances for ``All Races'' used in the models. Based on the genotype and cancer diagnosis statuses of each individual, we simulated results from biomarker testing for ER, CK5/6, CK14, PR, and HER for those with breast cancer and MSI for those with colorectal cancer. The code to simulate families and reproduce these simulation studies is publicly available \citep{liang2022panelprorepo}. 

\subsection{Performance Metrics}
\label{sec:diagnostic_metrics}
To evaluate the carrier probabilities from our models, we consider several performance metrics \citep{steyerberg2010assessing}. We use the area under the curve (AUC) as a measure for discrimination, the expected divided by the observed number of events (E/O) as a measure of calibration, and mean squared error (MSE) as a measure of overall accuracy. When computing these metrics, non-carriers are defined as individuals who are not variant carriers of any gene in the model. Percentile confidence intervals are obtained from 1000 bootstrap samples. 

\subsection{Back-Compatibility}
Four Mendelian risk prediction models are currently implemented in the BayesMendel R package (Table~3), each incorporating information from between 1-3 genes and cancers. Altogether, these models span seven genes (including a hypothetical PANC gene) and six cancers. 

To validate the reverse-compatibility of our approach, we fit these existing models using the PanelPRO and BayesMendel packages and evaluated their performance on 1,000,000 simulated families, using the same model parameters and settings. These families were simulated assuming only the presence of the gene-cancer associations incorporated in the four existing models (orange in Figure~1), and based on family structures sampled from the HCP cohort. Because PanelPRO and BRCAPRO from the BayesMendel package handle contralateral breast cancer using different approaches, we eliminated it from consideration when fitting the models, to obtain a more direct comparison. The resulting metrics and carrier probabilities are virtually identical between the two sets of models, as expected (Supplemental Figure~S1 and Supplemental Table~S1 \citep{liang2022panelprofigs}).

\subsection{PanelPRO-5BC}

We then simulated and evaluated 1,000,000 families under PanelPRO-5BC, a model that obtains carrier probabilities for ATM, BRCA1, BRCA2, CHEK2, and PALB2 based on family history of breast and ovarian cancer. As a point of reference, we also used the PanelPRO package to evaluate the simulated families with the BRCAPRO submodel, which only considers BRCA1 and BRCA2. Figure~2 plots metrics with bootstrap confidence intervals for PanelPRO-5BC; these metrics are also reported in Supplemental Table~S2 \citep{liang2022panelprofigs}.  

The results for the PanelPRO-5BC model and its BRCAPRO submodel are highly similar. 
The metrics for PanelPRO-5BC and BRCAPRO were also  similar to each other across bootstrap replicates, as described in Section~S7 of the supplementary material \citep{liang2022panelprofigs}. Incorporating simulated tumor marker testing for ER, PR, and HER in the model led to slightly better discrimination and precision of BRCA1 and BRCA2, as well as differences in calibration.


\subsection{PanelPRO-11}

As another illustration of the flexibility of our approach, we considered PanelPRO-11, an 11-gene (ATM, BRCA1, BRCA2, CDKN2A, CHEK2, EPCAM, MLH1, MSH2, MSH6, PALB2, and PMS2), 11-cancer (brain, breast, colorectal, endometrial, gastric, kidney, melanoma, ovarian, pancreatic, prostate, and small intestine) model. The performance metrics resulting from evaluating PanelPRO-11 on 1,000,000 simulated families are plotted in Figure~3 and are reported in detail in Supplemental Table~S3 \citep{liang2022panelprofigs}. We also evaluated the three BayesMendel submodels (BRCAPRO, MMRpro, Melapro) that are nested within PanelPRO-11. Including biomarker testing results can improve the metrics, particularly discrimination, for related genes such as BRCA1, BRCA2, MLH1, MSH2, MSH6, and PMS2. The most noticeable differences in discrimination are for MSH6 and PMS2. Incorporating tumor biomarker testing does not appear to meaningfully improve the discrimination for BRCA1, BRCA2, MLH2, and MSH2 likely because (unlike MSH6 and PMS2) these genes already have very high AUCs ($>0.9$ or even $>0.95$) in the baseline simulations. If the AUCs are close to the maximum of 1 to begin with, the range in which to improve after adding additional risk modifiers is very small. The very large calibration confidence intervals for EPCAM can be attributed to its low allele frequency, and the therefore low number of cases among the 1,000,000 simulated counselees. 

Throughout, the full PanelPRO-11 model's metrics are closely aligned with those of its submodels, with slightly better discrimination in some cases. In nearly all of the bootstrap samples, PanelPRO-11 has a better AUC and MSE. PanelPRO-11's improvement in terms of calibration is less consistent, but the differences in model calibration are typically small (less than 2\%) within bootstrap samples. Furthermore, whether or not PanelPRO improves over the submodel seems to be largely dependent on how well calibrated both models are; when both models are well-calibrated, PanelPRO's calibration will be better about half the time. See Section~S7 of the supplementary material for additional details \citep{liang2022panelprofigs}. PanelPRO-11 is thus able to produce the results of the existing BRCAPRO, MMRpro, and Melapro models, while also computing probabilities for being a carrier of ATM, CHEK2, EPCAM, PALB2, and PMS2.

The AUC for predicting pathogenic variant carriers of ``Any'' gene is 0.69. This is lower than the discrimination for most of the individual genes, many of which have AUCs above 0.8 or even 0.9. Groupings of genes that have individually high AUCs, like the BRCAPRO genes (BRCA1 and BRCA2) or the MMRpro genes (MLH1, MSH2, and MSH6), tend to retain good discrimination. However, including poorly-discriminating genes, as expected, pulls down the group AUC. This trend is observable for both PanelPRO-11 and the PanelPRO-5BC AUC results in Figure~2(a). 

Figure~4 plots the AUCs from this simulation study against the relative risks of developing each of the 11 cancers in PanelPRO-11 by age 70, for pathogenic variant carriers compared to noncarriers. Both male and female relative risks are shown; see Supplemental Figure~S2 for plots that separate between sexes \citep{liang2022panelprofigs}. Genes with lower AUCs correspond to those that have low relative risks for each of the eleven cancers. It appears that good discrimination for a given gene requires the gene to have a high relative risk for at least one cancer in the model. 
This trend is not expected for allele frequencies (Supplemental Figure~S3 \citep{liang2022panelprofigs}). Note that ATM and CHECK2 have both low penetrance and high prevalence, but their low AUC is likely attributable to the former.
The AUCs, allele frequencies, and relative risks plotted in Figure~4 and Supplemental Figures~S2-S3 are summarized in Supplemental Table~S4 \citep{liang2022panelprofigs}. 

We simulated an additional 1 million families generated by adding a latent PRS to the assumptions of PanelPRO-11. For the founders of a given family, we simulated 20 independent SNPs, each with minor allele frequency 0.1, and passed them down to their descendants based on Mendelian laws of inheritance. Each family member was then assigned a score calculated as the sum of the 20 SNPs in their genotype (1 being carrier and 0 being noncarrier). Each score was mapped to a penetrance-multiplying factor ranging between 0.8 and 1.2, with smaller scores corresponding to multiplying factors below 1 and larger scores corresponding to multiplying factors above 1. We then multiplied each individual's cancer penetrances (used for simulating cancer status) by the multiplying factor corresponding to their score, thereby modifying the simulated families' cancer risk in a heritable fashion that is not accounted for by PanelPRO. We also simulated a set of 1 million families using the same scheme, but with a score based on 40 independent SNPs that map to multiplying factors ranging between 0.6 and 1.4. 

The performance metrics for PanelPRO-11 (which does not account for the PRS) in these simulations with unmeasured genetic risk factors were virtually identical to those for the original simulation (Supplemental Figure~S4 and Table~S5 \citep{liang2022panelprofigs}). We designed the PRS simulations and the multiplying factors assigned to each score such that the expected number of cancer cases should remain similar to the original simulation. Counselees with the mean PRS will be assigned cancer statuses based on cancer penetrances that are the same as those used in the model. The cancer risk for many of the other counselees (i.e. those whose scores map to multiplying factors close to 1) will also be similar to the risk accounted for by the cancer penetrances in PanelPRO. So it is not entirely unexpected that PanelPRO's performance is largely unaffected when assessed by aggregate measures like the AUC.

\section{Data Application}
\label{sec:data_application}
\subsection{Model Validation}
For model validation, we applied PanelPRO-5BC and PanelPRO-11 to the data from the HCP study (Section \ref{sec:usc}). We also ran the BayesMendel models nested within these models (BRCAPRO for PanelPRO-5BC and BRCAPRO, MMRpro, and Melapro for PanelPRO-11). The code to reproduce the validation results is available in \cite{liang2022panelprorepo}. Some of the families in the HCP study report tumor marker testing results, so we ran the models with and without incorporating this extra risk-modifying information. To evaluate the models, we used the performance metrics described in Section \ref{sec:diagnostic_metrics} and obtained percentile confidence intervals from 1000 bootstrap samples. 

The metrics and confidence intervals for PanelPRO-5BC are reported in Figure~5 and Supplemental Table~S6; those for for PanelPRO-11 are reported in Figure~6 and Supplemental Table~S7 \citep{liang2022panelprofigs}. Metrics use outcome labels based on carrier status of individual genes as well as groups of genes. Both types remain fairly consistent between the full PanelPRO models and their submodels. The results for models run with risk modifiers are often similar to those without risk modifiers. The study includes a handful of noncarrier counselees for whom PanelPRO-11 greatly overpredicts the MSH6 carrier probability compared to MMRpro. Inspection of individual HCP pedigrees reveals that many of these counselees have no or limited family history of colorectal and endometrial cancer, but have family history for other cancers, such as brain, gastric, and ovarian cancer. Both models incorporate family history for colorectal and endometrial cancers, but only PanelPRO-11 incorporates associations between Lynch genes and these additional cancers, which are also associated with other genes. In this scenario, PanelPRO-11 points to inherited susceptibility, distributing probabilities across MLH1, MSH2, and MSH6 as well as other genes associated with these cancers. When we evaluate performance for Lynch syndrome genes MLH1, MSH2, and MSH6 only, PanelPRO slightly over-predicts on the overall average for these carrier probabilities compared to the MMRpro submodel, and suffers a slight worsening of the AUC for MSH6.

We acknowledge that the small sample size of the data-processed cohort makes it difficult to detect meaningful differences. The wide bootstrap confidence intervals are likely also attributable to the small number of cases. Similarly, the observation that incorporating risk modifiers into the models makes a limited impact on the results may be due to the relatively low number of families with tumor marker information. Section S7 of the supplementary material \citep{liang2022panelprofigs} contains discussion on how often PanelPRO-5BC and PanelPRO-11 improve over the BayesMendel submodels in the 1000 bootstrap replicates; the full model does not consistently have better metrics across bootstraps. These results are again likely due to limitations in the sample size and ascertainment process for the HCP data that cause it to deviate from the makeup of the general population, for which the model parameters were estimated. The HCP cohort used for data validation is highly ascertained, such that much of the family history in the pedigrees includes breast and colorectal cancers. Family history for additional cancers incorporated by PanelPRO beyond those in the syndrome-specific BRCAPRO and MMRpro is underrepresented compared to the general population. We would expect PanelPRO to return better predictions for families that have history of cancers and/or gene-cancer associations not incorporated in the existing syndrome-specific models.

The AUC confidence intervals for some genes, especially low-penetrant genes, overlap with 0.5. However, we note that the lower bound for the ``Any'' AUC is consistently above 0.5 for both PanelPRO-5BC and PanelPRO-11. Moreover, it is still beneficial to include lower-penetrant genes so that PanelPRO can estimate the probabilities of carrying pathogenic variants of these genes. 

A key contribution of the multi-syndrome predictions enabled by PanelPRO is the ability to acknowledge syndrome overlap in counseling individuals at risk. To illustrate, Figure~7 shows the probabilities of carrying a pathogenic variant of any PanelPRO-11 gene against the probabilities for carrying a pathogenic variant of any BRCAPRO gene (BRCA1 or BRCA2, left) and any MMRpro gene (MLH1, MSH2, or MSH6, right) in the HCP cohort. Of the 150 HCP counselees who are carriers of any pathogenic variant in PanelPRO-11, 61 had carrier probabilities above a 2.5\% threshold for PanelPRO but not BRCAPRO and 0 had carrier probabilities above 2.5\% for BRCAPRO but not PanelPRO. Similarly, 89 counselees had carrier probabilities above above 2.5\% for PanelPRO but not MMRpro, and 0 had carrier probabilities above 2.5\% for MMRpro but not PanelPRO. This demonstrates that the potential refinement in counseling practice by jointly considering syndromes whose phenotypes overlap can affect a large proportion of families. 

In Figure~S5 \citep{liang2022panelprofigs}, the points are now colored to represent the cancer types recorded in each counselee's pedigree. As expected, both PanelPRO-11 and the syndrome-specific models tend to correctly identify carriers with family history of only the cancers included in the syndrome-specific models. However, PanelPRO often has the advantage when the counselee has family history of both the cancers that are included in the syndrome-specific models as well as additional cancer types and especially when the counselee \textit{only} has other related cancers that are not included in the syndrome-specific model. 

\subsection{Sensitivity to Family Size}
To examine the performance that can be achieved by PanelPRO-5BC and PanelPRO-11 when very extensive family history is available, we simulated additional families with a large number of relatives. Each simulated counselee has four maternal aunts, maternal uncles, paternal aunts, paternal uncles, sisters, and brothers. The counselee and their siblings each have four daughters and four sons, for a total of 112 simulated relatives in each family. These are unrealistically large for most clinical applications, and serve here to provide a best case scenario.

Performance metrics and bootstrap confidence intervals obtained from evaluating PanelPRO-5BC and PanelPRO-11 on the HCP cohort, simulated families with structures sampled from the HCP cohort, and large simulated families are reported in Supplemental Figures~S10-S11 and Supplemental Tables~S12-S13 \citep{liang2022panelprofigs}. To make the simulation results more comparable to the results from the HCP data, we computed these metrics based on a subset of the simulated families that has the same number of families as the HCP cohort, with the same number of pathogenic variant carriers for each gene. While the metrics usually look best when PanelPRO can take advantage of a detailed family history, the results from the real data and a simulation with fewer relatives often do not lag too far behind. We also note that the results from the HCP cohort are typically better calibrated than those from either set of simulated families, albeit with wide confidence intervals. The process of subsetting the simulated families to have the same number of carriers as the HCP cohort does not match the ascertainment process for the cohort, which is likely the driving force behind this phenomenon.

\section{Discussion}
\label{sec:discussion}
We have developed PanelPRO, a general, computationally efficient framework for Mendelian risk models that incorporates an arbitrary number of genes and cancers or syndromes (groups of cancers that frequently occur together). Our model can leverage more family history information than its syndrome-specific counterparts by including multiple cancer types. It also allows for the development of models that address syndromes caused by potentially many hereditary genetic factors. Extending beyond syndrome-specific models, PanelPRO integrates multiple gene-cancer associations and is flexible enough to incorporate any number of additional associations as they may be discovered in the future. This more comprehensive approach for quantitative risk assessment can be used to make a better determination of which individuals should undergo genetic testing and targeted preventative interventions. 

By limiting the maximum number of pathogenic variants to consider in an individual's genotype, we keep the problem computationally feasible \citep{madsen2018efficient} even as the number of genes in the model grows, trading off between speed and accuracy. \cite{lee2021panelpro} describes the package in detail, provides practical examples for users, and contains further discussion on computational performance. For moderately-sized families, models with as many as 10 genes can be run within a few seconds, and models with as many as 20 genes can be run in under a minute, making this package suitable for clinical applications. As our simulation studies demonstrate, PanelPRO is fully reverse-compatible with existing models. These results suggests that the more general PanelPRO could easily be adopted by current users of the BayesMendel models. For simplicity and ease of use, some of the results from PanelPRO could be masked from viewers who are only interested in specific genes and cancers. 

Validation of models jointly predicting the carrier status of a large number of genes remains a challenge. The rich, multi-class nature of the labels, the relative rarity of pathogenic variants in certain genes, and the broad variety of gene panels that are emerging in clinical practice all contribute to the difficulty of assembling cohorts that can definitively establish performance. The HCP study considered here is among the best available, but nonetheless, the small sample size is not sufficient for addressing high resolution questions about the more rare genes. We do succeed in establishing the important conclusion that the five- and eleven-gene models can be implemented clinically without detriment of performance compared to current practice. While PanelPRO's ability to discriminate pathogenic variant carriers of any gene in the model is reduced as one incorporates genes with low penetrances, doing so can still be beneficial for identifying additional at-risk individuals. Furthermore, PanelPRO's carrier probabilities for any pathogenic variant detect a high number of true carriers that would have been missed by the syndrome-specific models. We hope that these results encourage the collection of more multi-gene panel testing data; as more suitable datasets become available, we plan to perform additional validation. 

The PanelPRO package \citep{lee2021panelpro} includes some functionalities that were not illustrated in our analysis. Besides tumor markers, our models currently consider prophylactic surgeries and germline testing as risk modifiers, and additional modifiers can be flexibly incorporated. Our framework for incorporating risk modifiers generalizes beyond surgical interventions and can also be used to support polygenic risk scores, assuming that suitable parameter estimates (modified penetrances, risk ratios, or hazard ratios) can be obtained. An observed PRS could simply be treated as an additional risk modifier. This approach can be applied to the counselee and any number of relatives who report PRSs. A further generalization could also explicitly include individual SNPs as additional genes, and model risk modification for each. While computationally challenging at the moment, particularly for a multi-syndrome model, this approach would account for additional genetic aggregation among individuals for whom the PRS is unobserved. When the cancer penetrance information is available, PanelPRO can return carrier probabilities for genotypes with a given pathogenic variant on both alleles as well as one allele. A future package release will handle multiple gene variants for a given gene. 

We made several modeling assumptions, such as assuming that genotypes are independent across first-generation family members, followed by conditionally independent genotypes for second-generation family members and so on for subsequent generations \citep{berry1997probability}. Including many genes in model therefore necessitates relying more heavily on these conditional independence assumptions. More importantly, we also assumed that the phenotypes are independent conditional on the individual's genotype. Censoring and deaths by causes unrelated to a given cancer are taken to be independent of the time to cancer diagnosis and non-informative with respect to carrier status. We assumed an absence of competing risks, but as the number of cancers included in the model increases, many of the competing risks will naturally be accounted for. PanelPRO is further limited by uncertainties in its inputs. The model's ability to estimate carrier probabilities based on family history depends in part on the input pedigree, which may contain errors or missing information \citep{braun2014extending}. The model also relies on robust estimation of the model parameters--- namely, the allele frequencies and penetrances, as well as parameters for secondary cancers and risk modifiers--- based on existing literature. As the number of genes, cancers, and associations grows, more model parameters will need to be obtained and carefully evaluated to ensure that they improve model performance overall. Future versions of PanelPRO may also explicitly incorporate uncertainty in the input parameters, for example via the Monte-Carlo approach in \cite{parmigiani1998determining}.

We would generally expect the inclusion of cancers that are associated with many genetic variants to provide more gains than cancers that are associated with only one or two, but the number of associations is not the only consideration. There is also the size/strength of the association (e.g. a highly penetrant vs. moderately penetrant gene) as well as the prevalence of deleterious variants and the rate of the cancer in the non-carrier population. Larger families and families with more complete information are more informative and should lead to more accurate predictions by PanelPRO (as well as other models). The probability provided by Mendelian models of the kind described here can be thought of as integrating over all unknowns (genotypes and phenotypes) of any pedigree that is a superset of the observed pedigree. This justifies looking at calibration across varying family sizes. However, this result relies on model assumptions whose degree of realism may begin to deteriorate as the family size increases \citep{huang2021variation}. 

Despite the numerous practical challenges remaining in this area, we hope that the availability of our comprehensive framework and associated open source software will contribute to a more systematic consideration of the relations between inherited susceptibility and cancer phenotypes, support more systematic prevention efforts, decrease barriers to interdisciplinary management of families at high risk for cancer, and encourage the adoption of general-purpose early detection strategies \citep{Fiala2019jalm,Cristiano2019n} among relatives of carriers.

\section*{Acknowledgments}
\label{sec:acknowledgements}

We gratefully acknowledge support from the National Cancer Institute at the National Institutes of Health grants 5T32CA009337 (JWL) and 4P30CA006516 (GP).

\bibliographystyle{apacite}
\bibliography{bibliography}

\section*{Supporting Information}
\label{sec:supporting}
Supplemental figures, tables, and discussion can be viewed in the separate appendix file \citep{liang2022panelprofigs}. The PanelPRO R package is freely available for research purposes at \url{https://projects.iq.harvard.edu/bayesmendel/panelpro} \citep{lee2021panelpro}. Code to perform the analysis and 
generate the figures in this paper is available at \url{https://github.com/janewliang/PanelRePROducible} \citep{liang2022panelprorepo}.

\end{document}